\documentstyle[prl,aps,twocolumn]{revtex}

\begin{document}

\title{
{\normalsize\begin{flushright}
UB-ECM-PF-98/13\\
ITP-UH-11/98\\
hep-th/9807113\\[1.5ex]
\end{flushright}}
Infinitely many rigid symmetries of kappa-invariant D-string actions
}
\author{
Friedemann Brandt\,$^a$, Joaquim Gomis\,$^b$, 
David Mateos\,$^b$, Joan Sim\'on\,$^b$}

\address{
$^a$ Institut f\"ur Theoretische Physik, Universit\"at Hannover,
Appelstra\ss e 2, D--30167 Hannover, Germany\\
$^b$ Departament ECM,
Facultat de F\'{\i}sica,
Universitat de Barcelona and 
Institut de F\'{\i}sica d'Altes Energies,
Diagonal 647,
E-08028 Barcelona, Spain
\\[1.5ex]
\begin{minipage}{14cm}\rm\quad
We show that each rigid symmetry of a D-string action is contained in a 
family of infinitely many symmetries. In particular, kappa-invariant 
D-string actions have infinitely many supersymmetries. The result is not 
restricted to standard D-string actions, but holds for any two-dimensional 
action depending on an abelian world-sheet gauge field only via the field 
strength. It applies thus also to manifestly $SL(2,Z)$ covariant D-string 
actions. Furthermore, it extends analogously to $d$-dimensional actions with 
$(d-1)$-form gauge potentials, such as brane actions with dynamical tension.
\\[1ex]
PACS numbers: 11.25.-w, 11.25.Hf, 11.30.-j, 11.30.Pb\\
Keywords: D-string, supersymmetry, kappa invariance, brane actions
\end{minipage}
}

\maketitle

\section*{Introduction and conclusion}

In \cite{paper1,paper2,paper3} a complete classification of the
rigid symmetries of bosonic D-string actions was given and several
examples were worked out explicitly, both in flat and curved
backgrounds. In particular it  was shown that each rigid symmetry 
is contained in a family of infinitely many rigid symmetries.
All these families together form a loop (or loop-like) symmetry algebra.
Two important examples were those of a D-string in the near 
horizon geometries of D3 and D1+D5 branes \cite{paper3}. 
The near horizon metrics involve
$AdS$ factors whose isometry groups are $SO(2,4)$ and
$SO(2,2)$ respectively,
and thus
the symmetries of
the D-string action in these backgrounds contain an infinite
loop generalization of these conformal symmetries.

In this paper we show that the above structures extend to supersymmetric 
and, in particular, to kappa-invariant D-string actions.
Hence, these actions have actually infinitely many supersymmetries,
forming infinite dimensional loop-generalizations of the familiar 
supersymmetry algebras (in flat or curved backgrounds).

As in the purely bosonic case, the infinite symmetry structure
is a direct consequence of the presence of the Born-Infeld gauge field 
$A_\mu$. This will become particularly clear from the way in which we shall
derive the result. Namely we shall use a simple general
argument which neither makes use of the particular form of the
action nor of any specific properties of the target space or its symmetries.
Rather, the argument uses solely that the Lagrangian depends on $A_\mu$
only via the field strength 
$F_{\mu\nu}=\partial_\mu A_\nu-\partial_\nu A_\mu$, and that the
world-volume is two-dimensional. 

Our result is thus not restricted to D-string actions of the
Born-Infeld type but applies actually to a much larger class 
of two dimensional 
actions containing a $U(1)$ gauge field. Moreover, if such an action
contains several $U(1)$ gauge fields only via their field strengths, 
then the argument applies to each of these gauge fields separately, 
yielding an even larger symmetry structure. 
In particular this applies to the manifestly $SL(2,Z)$-covariant D-string 
actions constructed in \cite{PKT,CT} which contain two $U(1)$ gauge fields.

Furthermore we shall show that the argument is actually not restricted to
two-dimensional actions. Rather, it extends analogously to
$d$-dimensional actions containing $(d-1)$-form gauge potentials 
only via their (abelian) field strengths. In the context of branes,
such actions have been discussed in \cite{PKT,CT,PKT2,BLT,CW,BT,BST}
where the $(d-1)$-form gauge potentials serve to implement the brane 
tension dynamically (as an integration constant). 
The two-dimensional 
case appears to be somewhat special in the context of D-branes, as only 
in this case the Born-Infeld gauge field itself 
serves as a $(d-1)$-form gauge potential. 

Therefore the paper focusses mainly on the existence and construction 
of infinite families of symmetries of D-string actions.
We do not provide a complete characterization of all these families 
of symmetries. From the
results in the bosonic case, we expect that such a characterization 
can be given in terms of generalized super-Killing vector equations.
Here we just remark that the families of symmetries of D-string actions do not 
necessarily correspond one-to-one to the target space (super-) isometries.
For instance, in the bosonic case there are backgrounds which
admit the presence of dilatational symmetries in addition to families of
symmetries arising from the target space isometries
\cite{paper1,paper2,paper3}. We shall provide a supersymmetric
version of these dilatational symmetries in a flat background 
which however does not seem to extend (at least not straightforwardly)
to the kappa-invariant case 
as the Wess-Zumino term breaks these dilatational symmetries.

One interesting case would be that of the kappa-invariant
D-string action in a D1+D5 supersymmetric background, which
could be constructed along the lines of \cite{kallosh,Met.Tsey}.
It follows from our results that such an action should contain 
among its rigid symmetries an infinite loop generalization
of the background isometry supergroup  
$SU(1,1\mid 2) \times SU(1,1\mid 2)$.

Finally, we wish to stress that the nature of the 
infinite symmetry structure
described here differs from the infinite conformal symmetry
of gauge fixed two dimensional sigma models discussed 
in \cite{hull,WN}.
Namely, these conformal symmetries of sigma models are a mixture of
finitely many target space symmetries and infinitely many (conformal)
world-sheet symmetries which arise as residual symmetries 
from world-sheet diffeomorphisms in appropriate gauges of the latter.
In contrast, the infinitely many symmetries
of D-string actions discussed here exist in addition to
the world-sheet diffeomorphisms and are thus present even before
gauge fixing the latter.

\section*{The general argument in the two-dimensional case}

We consider a two-dimensional action
$S=\int d^2\sigma L$ with a Lagrangian $L$ which depends
on the gauge field $A_\mu$ only via its field strength 
$F_{\mu\nu}=\partial_\mu A_\nu-\partial_\nu A_\mu$ and, possibly,
derivatives thereof (or which
can be brought into such a form by means of a partial integration). 
We shall denote by $\{Z^M\}$ all the other fields which occur in the action.
For instance, in the case of standard bosonic D-string actions 
$\{Z^M\}$ contains only the target space
coordinates $x^m$, while it contains in addition the fermionic fields
$\theta^\alpha$ in the supersymmetric or kappa-invariant case.
If the action contains additional abelian gauge fields (as, e.g., in
\cite{PKT,CT}), the latter count also among the $Z^M$, and $A_\mu$
can be any of those gauge fields which enter the action only via
their field strengths.

As $L$ depends by assumption on $A_\mu$ only via $F_{\mu\nu}$ and 
derivatives thereof,
its Euler Lagrange derivative ${\hat\partial L}/{\hat\partial A_\mu}$ with
respect to $A_\mu$ takes the form
\begin{equation}
\frac{\hat\partial L}{\hat\partial A_\mu}=
\epsilon^{\mu\nu}\partial_\nu \varphi
\label{GA1}
\end{equation}
where $\varphi=\epsilon_{\nu\mu} \,
\partial L/\partial F_{\mu\nu}$ 
(using $\epsilon^{01}=\epsilon_{10}=1$).
Note that we have used here that we are dealing with a two
dimensional theory, as we took advantage of the fact that
$F_{\mu\nu}$ is proportional to $\epsilon_{\mu\nu}$.

We shall now show that any rigid symmetry of $S$ is actually contained
in a family of infinitely many rigid symmetries. Let us therefore
assume that there are infinitesimal transformations $\Delta Z^M$
and $\Delta A_\mu$ which generate a symmetry of the action,
i.e., by assumption the $\Delta$-variation of the Lagrangian 
is a total derivative, $\Delta L=\partial_\mu k^\mu$. 
This invariance property is equivalent to
\begin{equation}
(\Delta Z^M)\,\frac{\hat\partial L}{\hat\partial Z^M}+
(\Delta A_\mu)\,\frac{\hat\partial L}{\hat\partial A_\mu}
=\partial_\mu j^\mu_\Delta\ .
\label{GA3}
\end{equation}
Here $j^\mu_\Delta$ is of course nothing but the Noether current 
associated with $\Delta$. We claim that the
following transformations $\tilde \Delta$ generate further
rigid symmetries of the action,
\begin{eqnarray}
\tilde\Delta Z^M&=& \lambda(\varphi)\, \Delta Z^M
\label{GA4}\\
\tilde\Delta A_\mu &=&  \lambda(\varphi)\, \Delta A_\mu
-\frac{d\lambda(\varphi)}{d \varphi}\,
\epsilon_{\mu\nu} j^\nu_\Delta
\label{GA5}
\end{eqnarray}
where $\lambda(\varphi)$ is an arbitrary function of the quantity
$\varphi$ occurring in Eq.\ (\ref{GA1}).
Indeed, using Eqs.\ (\ref{GA1}) and (\ref{GA3}) one easily verifies that
\begin{equation}
(\tilde\Delta Z^M)\,\frac{\hat\partial L}{\hat\partial Z^M}+
(\tilde\Delta A_\mu)\,\frac{\hat\partial L}{\hat\partial A_\mu}
=\partial_\mu \left[\lambda(\varphi)\, j^\mu_\Delta\right]\ .
\label{GA6}
\end{equation}
This implies $\tilde\Delta L=\partial_\mu \tilde k^\mu$ and thus 
$\tilde\Delta$ generates a symmetry of the action.
Furthermore Eq.\ (\ref{GA6}) shows that the Noether current
associated with $\tilde\Delta$ arises from the one associated
with $\Delta$ simply through multiplication with $\lambda(\varphi)$,
\begin{equation}
j^\mu_{\tilde\Delta}=\lambda(\varphi)\, j^\mu_\Delta\ .
\label{GA7}
\end{equation}
Hence, given a symmetry $\Delta$ of the action, any choice
$\lambda(\varphi)$ yields another symmetry $\tilde\Delta$,
and thus gives indeed rise to a family of infinitely many symmetries.
Notice that if $\Delta$ is a linear combination of a set of
independent rigid symmetries, $\Delta = \epsilon^i \Delta_i$, each
of the symmetries $\Delta_i$ yields a corresponding family
of symmetries $\tilde \Delta_i$
through functions $\lambda^i(\varphi)$.

\section*{Kappa invariant D-string}

As a first example of the above statement, we consider the 
kappa-invariant D-string action in a flat ten-dimensional background 
with two target space Majorana-Weyl fermions 
$\theta_1^\alpha,\theta_2^\alpha$ of the same chirality 
(type IIB case). Using the notation and 
conventions of \cite{schwarz,kiyoshi} (in particular
$\theta=\theta_1+\theta_2$ with
$\theta_1=\frac 12(1+\tau_3)\theta$
and $\theta_2=\frac 12(1-\tau_3)\theta$), the action reads
\begin{eqnarray}
S & = & -T\int d^2\sigma\, 
\sqrt{-\det({\cal G}_{\mu\nu} + {\cal F}_{\mu\nu})} +
T\int \Omega_{(2)}(\tau_1)
\label{action}
\end{eqnarray}
where
\begin{eqnarray}
& {\cal G}_{\mu\nu}= \Pi^m_{\mu}\Pi^n_{\nu}\eta_{mn} \quad , \quad
\Pi^m_{\mu} = \partial_{\mu}X^m - \bar \theta\Gamma^m\partial_{\mu}\theta &
\nonumber \\
& {\cal F} = d\,A - \Omega_{(2)}(\tau_3) &
\nonumber \\
& \Omega_{(2)}(\tau_i) = -\bar \theta\Gamma_m\tau_id\,\theta(
d\,x^m + \frac{1}{2}\bar \theta\Gamma^md\,\theta). &
\label{def1}
\end{eqnarray}
The above action is known to be invariant up to a total derivative
under super-Poincar\'e transformations 
$a^m \Delta_m + \frac{1}{2}a^{mn}\Delta_{mn}
+ \epsilon^{\alpha}\Delta_\alpha$,
where $a^m$, $a^{mn}=-a^{nm}$ and $\epsilon^{\alpha}
=\epsilon_1^{\alpha}+\epsilon_2^{\alpha}$ 
are constant infinitesimal parameters associated with Poincar\'e and 
supersymmetry transformations,
respectively, while $\Delta_m$, $\Delta_{mn}$, $\Delta_\alpha$
are the corresponding generators. They act as follows
\begin{eqnarray}
& \Delta_n x^m = \delta^m_n \quad , \quad \Delta_n \theta^\alpha=
\Delta_n A_\mu = 0 & \label{traf0a}\\
& \Delta_{pq} x^m = (\delta^m_p \eta_{qr} - \delta^m_q \eta_{pr})x^r
\ , \ \Delta_{pq}\theta^\alpha = \frac{1}{2}
(\Gamma_{pq}\theta)^\alpha & \nonumber \\
& \Delta_{pq} A_\mu = 0 & \label{traf0b}\\ 
& \Delta_\beta x^m = (\bar \theta\Gamma^m)_\beta \quad , \quad
\Delta_\beta \theta^\alpha = \delta^\alpha_\beta & \nonumber \\
& \Delta_\alpha  A_\mu = (\bar \theta\tau_3\Gamma_m)_\alpha\partial_\mu x^m &
\nonumber \\
& - \frac{1}{6}\left[
(\bar \theta\tau_3\Gamma_m)_\alpha\bar \theta\Gamma^m \partial_\mu\theta + 
(\bar \theta\Gamma_m)_\alpha
\bar \theta\tau_3\Gamma^m \partial_\mu\theta)\right] &
\label{traf0}
\end{eqnarray}

Up to the irrelevant factor $T$, (\ref{GA1}) yields
in this case
\begin{equation}
\varphi = \frac{\bar \varphi}{\sqrt{1 - \bar \varphi^2}}
\quad , \quad 
\bar \varphi = 
\frac{\epsilon^{\mu\nu}{\cal F}_{\mu\nu}}{2\sqrt{-{\cal G}}}
\label{scalar}
\end{equation}
where ${\cal G}=\det({\cal G}_{\mu\nu})$.
It is now straightforward to apply Eqs.\ (\ref{GA4},\ref{GA5})
to any $\Delta \in \{\Delta_m\, , \Delta_{mn}\, , \Delta_\alpha\}$,
using (\ref{traf0a}--\ref{traf0}) and the corresponding
Noether currents. The latter are given by
\begin{eqnarray}
j^\mu_{\Delta_m} &=& \hat\Pi^\mu_m
-\epsilon^{\mu\nu}\bar\theta\hat\Gamma_m\partial_\nu\theta \\
j^\mu_{\Delta_{mn}} &=& 
\hat\Pi^\mu_p(2\delta^p_{[m}\eta_{n]q}x^q
-\frac 12\bar\theta\Gamma^p\Gamma_{mn}\theta)
\nonumber\\
& &-\epsilon^{\mu\nu}\bar\theta\hat\Gamma_p\partial_\nu\theta
(2\delta^p_{[m}\eta_{n]q}x^q
-\frac 14\bar\theta\Gamma^p\Gamma_{mn}\theta)
\nonumber\\
& &+\frac 12\epsilon^{\mu\nu}\bar\theta\hat\Gamma_p\Gamma_{mn}\theta
(\partial_\nu x^p-\frac 12\bar\theta\Gamma^p\partial_\nu\theta)
\\
j^\mu_{\Delta_{\alpha}} &=&
(\bar\theta\Gamma^m)_\alpha(2\hat\Pi^\mu_m
-\frac 43\epsilon^{\mu\nu}\bar\theta\hat\Gamma_m\partial_\nu\theta)
\nonumber\\
& & 
-\epsilon^{\mu\nu}(\bar\theta\hat\Gamma_m)_\alpha
(2\partial_\nu x^m-\frac 23\bar\theta\Gamma^m\partial_\nu\theta)
\label{traf1}
\end{eqnarray}
where
\begin{eqnarray}
\hat\Pi^\mu_m &=& \sqrt{-{\cal G}(1+\varphi^2)}\,{\cal G}^{\mu\nu}
\eta_{mn}\Pi^n_\nu
\\
\hat\Gamma_m&=&\Gamma_m(\varphi\tau_3-\tau_1).
\end{eqnarray}
Notice that each $\Delta$ has its corresponding arbitrary function
$\lambda(\varphi)$, which can be expanded in an appropiate basis
(e.g.\ in powers of $\varphi$)
to get the loop version of the corresponding super-Poincar\'e
algebra, cf.\ \cite{paper1,paper3}.

\section*{Purely supersymmetric D-string}

By purely supersymmetric D-string, we mean a supersymmetric 
D-string
with no coupling to the RR-potentials and 
NS-NS two form. Again we consider an action in a flat background,
\begin{equation}
S= -T\int d^2\sigma \,\sqrt{-\det({\cal G}_{\mu\nu} + F_{\mu\nu})}
\end{equation}
with ${\cal G}_{\mu\nu}$ as in (\ref{def1}) and
$F_{\mu\nu}=\partial_\mu A_\nu-\partial_\nu A_\mu$.
This example illustrates that Born-Infeld
type actions can have more symmetries than those associated
with background (super-) isometries, as was already pointed out 
in~\cite{paper1,paper3}. Namely, in addition to the
super-Poincar\'e symmetries%
\footnote{In this case, the gauge field does not transform at all under
the super-Poincar\'e transformations due to the non-appearance of the
NS-NS two form (in particular, $\Delta_\alpha A_\mu = 0$). 
Furthermore, in contrast to the kappa symmetric case,
the Lagrangian itself is exactly supersymmetric,
not just up to a total derivative.},
the action has a dilatational invariance generated by
\begin{eqnarray}
& \Delta_d \,x^m = x^m \quad , \quad \Delta_d \,\theta = \frac{1}{2}\,
\theta & \\
& \Delta_d \,A_\mu = 2 (1-\varphi^{-2})A_\mu &
\end{eqnarray}
where 
\begin{equation}
\varphi = \frac{\bar \varphi}{\sqrt{1 - \bar \varphi^2}}
\quad , \quad 
\bar\varphi= \frac{\epsilon^{\mu\nu} F_{\mu\nu}}
{2\sqrt{-{\cal G}}}\ .
\end{equation}
Indeed, the Lagrangian is invariant under $\Delta_d$
up to a total derivative,
\begin{equation}
\Delta_d\, L=\partial_\mu (4\varphi^{-1}A_\nu \epsilon^{\nu\mu}).
\end{equation}
The 
corresponding symmetries (\ref{GA4},\ref{GA5}) read as follows 
\begin{eqnarray}
& \tilde\Delta_d\, x^m = \lambda(\varphi)\,x^m \quad, \quad 
\tilde\Delta_d\, \theta = 
\frac 12\, \lambda(\varphi)\,
\theta & \\
& \tilde\Delta_d A_\mu = 2\lambda(\varphi)(1-\varphi^{-2})A_\mu +
\frac{d\lambda(\varphi)}{d\varphi}
\{2(\varphi+\varphi^{-1})A_\mu 
& \nonumber \\
& -\sqrt{-{\cal G}(1+\varphi^2)}\,\epsilon_{\mu\nu}
{\cal G}^{\nu\varrho}\Pi^n_\varrho \eta_{nm}
[x^m -\frac{1}{2}\bar \theta\Gamma^m
\theta]\}. &
\end{eqnarray}

\section*{The general argument in higher dimensions}

The argument given in the two-dimensional case can
be easily generalized to higher dimensions. Consider
a $(p+1)$-dimensional action $S= \int d^{p+1} \sigma L$
which depends on a $p$-form gauge field $A_{\mu_1 \ldots 
\mu_p}$ only through its field strength 
$F_{\mu_1\mu_2 \ldots \mu_{p+1}} = (p+1)\partial_{[ \mu_1}
A_{\mu_2 \ldots \mu_{p+1} ]}$ and derivatives thereof. 
Again we denote by 
$\{Z^M\}$ all other fields in the action. Similarly 
to the two-dimensional case, the Euler-Lagrange derivatives
of $L$ with respect to $A_{(p)}$ are then
\begin{equation}
\frac{\hat\partial L}{\hat\partial A_{\mu_1 \ldots \mu_p}} =
\epsilon^{\mu_1 \ldots \mu_p\nu} \, \partial_\nu\varphi\ .
\label{GGA1}
\end{equation}
Note that in this case we have used that $F_{\mu_1 \ldots \mu_{p+1}}$
is proportional to $\epsilon_{\mu_1 \ldots \mu_{p+1}}$, since we
are considering a $(p+1)$-dimensional theory.

In order to show that any rigid symmetry of $S$ is 
contained in a family of infinitely many rigid symmetries,
we proceed along the lines of the two-dimensional case.
Thus, let $\Delta Z^M$ and $\Delta A_{\mu_1 \ldots \mu_p}$
be infinitesimal transformations which generate a symmetry 
of $S$. This means that
\begin{equation}
(\Delta Z^M)\,\frac{\hat\partial L}{\hat\partial Z^M}+
(\Delta A_{\mu_1 \ldots \mu_p})\,\frac{\hat\partial L}{\hat\partial  
A_{\mu_1 \ldots \mu_p} }
\, =\,\partial_\mu j^\mu_\Delta\ .
\label{GGA2}
\end{equation}
If this is so, then the transformations
\begin{eqnarray}
\tilde\Delta Z^M&=& \lambda(\varphi)\, \Delta Z^M \\
\tilde\Delta A_{\mu_1 \ldots \mu_p} &=&  \lambda(\varphi)\, 
\Delta A_{\mu_1 \ldots \mu_p} \\
&&- \frac{1}{p!} \,
\frac{d\lambda(\varphi)}{d \varphi}\,
\epsilon_{\mu_1 \ldots \mu_p \nu} j^\nu_\Delta
\end{eqnarray}
where $\lambda(\varphi)$ is an arbitrary function of the quantity
$\varphi$ that appears in Eq. (\ref{GGA1}), also generate 
rigid symmetries of the action (we have used
$\epsilon^{01 \ldots p} = -\epsilon_{01 \ldots p} = 1$). 
Indeed, making use of Eqs. 
(\ref{GGA1}) and (\ref{GGA2}) it is easily checked that
\begin{equation}
(\tilde\Delta Z^M)\,\frac{\hat\partial L}{\hat\partial Z^M}+
(\tilde\Delta A_{\mu_1 \ldots \mu_p})\,\frac{\hat\partial L}{\hat\partial  
A_{\mu_1 \ldots \mu_p} }
\, =\,\partial_\mu [ \lambda(\varphi) \, j^\mu_\Delta ]
\end{equation}
This shows that $\tilde\Delta$ generates a symmetry of the action,
and also that the associated Noether current is simply 
$j^\mu_{\tilde\Delta}= \lambda(\varphi) \, j^\mu_\Delta$.

\section*{Examples: D-Branes and M-Branes}

The previous argument for the occurrence of infinite families
of rigid symmetries for $(p+1)$-dimensional actions depending
on $p$-form gauge potentials $A_{(p)}$ only through their 
field strengths $G_{(p+1)}= dA_{(p)}$ applies readily to different brane
actions. We will consider (super)D-branes and (super)M-branes. 
Both types of objects can be described by Lagrangian
densities in which the tension of the brane
is generated dynamically as an integration constant
of the field equations for the $p$-form gauge potential.
Their form is \cite{BT,BST}, see also \cite{CW},
\begin{equation}
L = \frac{1}{2v} \, \left[ L^2_K + (*G_{(p+1)})^2 \right]
\label{EX1}
\end{equation}
where $v$ is an indepependent worldvolume density and $*$
denotes the worldvolume Hodge dual. For instance, for a 
Dp-brane in a general $D=10$ supergravity background
one has \cite{BST} 
\begin{eqnarray}
L^2_K &=& e^{-2 \phi} \, \det (g_{\mu\nu} + \cal{F}_{\mu\nu}) \\
{\cal{F}}&=& dV - B \\
G_{(p+1)}&=& d A_{(p)} - C \, e^{\cal{F}},\quad
C=\oplus_k C_k
\end{eqnarray}
where $g$ is the induced metric, $V$ is the Born-Infeld
gauge field, $B$ is the pull-back of the NS-NS two-form
and $C_k$ are the pull-backs of the R-R gauge potentials. 
The corresponding expressions for the M2-brane and the M5-brane
in a $D=11$ supergravity background 
can be found in \cite{BLT} and \cite{BST} respectively.

In all these cases (for $p>1$) the `source' of an infinite
number of symmetries of the action is the worldvolume
$p$-form gauge potential $A_{(p)}$. The quantity $\varphi$
occurring in Eq. (\ref{GGA1}) in these cases takes the form
\begin{equation}
\varphi \propto \frac{*G_{(p+1)}}{v}\ .
\end{equation}
For every rigid symmetry $\Delta$ of these actions, the
construction explained in the previous section yields
an infinite family $\{\tilde\Delta\}$ of symmetries in
which the original one is included.
For instance, this applies to all (super)isometries
of the supergravity background, which were
shown in \cite{BT,BST} to yield
rigid symmetries of the corresponding 
brane action. 
It also applies to space-time scale transformations
under which (\ref{EX1}) in a flat background is invariant.
For the particular case of Dp-branes they take the form 
\footnote{For p-branes, this set of
transformations has already been written in ~\cite{BLT}.}:
\begin{eqnarray}
& x^m \; \longrightarrow \; k \, x^m & \nonumber \\
& \theta \; \longrightarrow \; k^{1/2} \, \theta & \nonumber \\
& V  \; \longrightarrow \; k^2 \, V & \nonumber \\
& A_{(p)}  \; \longrightarrow \; k^{p+1} \, A_{(p)} & \nonumber \\
& v \; \longrightarrow \; k^{2(p+1)} \, v &
\end{eqnarray}
The corresponding transformations for the M5-brane are obtained 
from the previous ones by setting $p=5$ and replacing the abelian one
form gauge potential $V$ by a self-dual two form $V_{(2)}^+$ 
that transforms with weight three, i.e. 
$V_{(2)}^+ \; \rightarrow \; k^3 \, V_{(2)}^+$.
 
The above discussion does not imply
that the actions for these branes in their `usual' form, i.e.,
without the fields $v$ and $A_{(p)}$,
have infinitely many rigid symmetries too. 
The reason is that $A_{(p)}$ cannot be eliminated
algebraically from the action. Rather, one eliminates it 
by solving
its field equation through an integration constant%
\footnote{Therefore the `usual' actions do not
arise from the Lagrangian (\ref{EX1}) simply
by substituting a solution to the field equations of $A_{(p)}$. 
Rather, before doing so, a term proportional to 
$*dA_{(p)}$ must be added to the Lagrangian (see \cite{PKT2}).}.
Hence, $\varphi$ turns into a constant once $A_{(p)}$ has been
eliminated in this manner. Accordingly, after eliminating 
$A_{(p)}$, $\Delta$ and $\tilde\Delta$
are not independent symmetries anymore, but
simply proportional to one another (so are the corresponding
Noether currents). In contrast, in
the two-dimensional (D-string) case
the `source' of the infinite number of 
symmetries is the Born-Infeld gauge field itself%
\footnote{Note that in the case of the $SL(2,Z)$-covariant
formulation \cite{PKT,CT} of the IIB superstring,
both sources of infinite symmetry are present, 
since the action contains two $U(1)$ gauge fields, one of
which can be considered as auxiliary and the other 
one as the Born-Infeld field.}.
Of course, the argument above does not disprove
the existence of an infinite set of symmetries for
D$p$-branes ($p>1$) and M-branes. This is an issue
which remains open.

\vspace{10mm}
{\em Acknowledgements.}
This work was supported in part by 
AEN95-0590 (CICYT), GRQ93-1047 (CIRIT) and
by the Commission of European Communities CHRX93-0362(04).
F.B. was supported by the Deut\-sche For\-schungs\-ge\-mein\-schaft.
D.M. is supported by a fellowship from
Comissionat per a Universitats i Recerca de la Generalitat de
Catalunya.

\end{document}